\documentclass[reprint,superscriptaddress,preprintnumbers,nofootinbib,nobibnotes,amsmath,amssymb,aps,prd,floatfix]{revtex4-2}

\usepackage{amsbsy,esint,mathrsfs,commath,amsfonts,mathtools,amsthm}
\usepackage[usenames,dvipsnames]{xcolor}
\usepackage{braket}
\usepackage{graphicx} 
\usepackage{hyperref}
\usepackage[caption=false]{subfig}

\hypersetup{
    colorlinks=true,
    linkcolor=Blue,
    filecolor=magenta,      
    urlcolor=Blue,
    citecolor=Blue,
    }

\newcommand*\diff{\mathop{}\!\mathrm{d}}

\newcommand{\Cred}{\breve C}
\newcommand{\Hred}{\breve H}
\newcommand{\Credg}{\Cred}
\newcommand{\Hredg}{\Hred}
\newcommand{\la}{\langle}
\newcommand{\ra}{\rangle}
\newcommand{\area}{A}

\begin{document}
 
\title{From loop quantum gravity to cosmology: the two-vertex model}

 \date{April 9, 2024}
 
 \author{\'Alvaro Cendal}
 \email{alvaro27@ucm.es}
 \affiliation{Departamento de F\'isica Te\'orica and IPARCOS, Universidad Complutense de Madrid, 28040 Madrid, Spain}  
 \author{I\~naki Garay} 
 \email{inaki.garay@ehu.eus}
 \affiliation{Department of Physics and EHU Quantum Center, University\,of\,the\,Basque\,Country  UPV/EHU, Barrio Sarriena s/n, 48940, Leioa, Spain}  
 \author{Luis J. Garay} 
 \email{luisj.garay@ucm.es}
 \affiliation{Departamento de F\'isica Te\'orica and IPARCOS, Universidad Complutense de Madrid, 28040 Madrid, Spain}  
\begin{abstract}
We study the notion of volume and its dynamics in the loop-quantum-gravity truncation known as the two-vertex model. We also show that its U$(N)$-symmetry reduction provides the old effective dynamics of loop quantum cosmology with an arbitrary perfect barotropic fluid content. A suitable modification of the Poisson bracket structure of the U($N$)-symmetric model leads to the loop quantum cosmology improved dynamics.
\end{abstract}

\preprint{IPARCOS-UCM-24-020}
\maketitle

\tableofcontents

\section{Introduction}
 
Loop quantum gravity (LQG)~\cite{thiemannModernCanonicalQuantum2008,rovellibook} proposes a theoretical framework to describe the geometry of spacetime at the Planck scale. This is particularly relevant in the vicinity of black-hole and big-bang general relativistic singularities.
A basis of the Hilbert space of LQG is formed by the so-called spin networks: SU(2) wave functions defined on graphs, labeled by spins on edges and intertwiners on   vertices. These spin networks are eigenstates of the area and volume operators, 
which have discrete spectra~\cite{Rovelli_Smolin_DiscretenessAreaVolume1995, Ashtekar_Lewandowski_QuantumTheoryGravity1997}.

The quantization techniques used in LQG have been mimicked on highly symmetric universes, giving rise to loop quantum cosmology (LQC)~\cite{Ashtekar_Singh_LoopQuantumCosmology2011, bojowald2008loop,agulloLoopQuantumCosmology2016}. Since its beginning, LQC has obtained several successes, describing different cosmological models~\cite{Ashtekar_Pawlowski_Singh_QuantumNatureBig2006, Ashtekar_Pawlowski_QuantumNature2006,Ashtekar_Pawlowski_Singh_Vandersloot_LoopQuantumCosmology2007, Martin-Benito_MenaMarugan_Pawlowski_LoopQuantizationVacuum2008} and physical predictions. The most relevant result of LQC is the prediction of a big bounce that avoids the initial singularity of the big bang~\cite{Ashtekar_Pawlowski_Singh_QuantumNatureBig2006}. 

Despite the advances made on both full LQG and LQC, the implementation of the dynamics and the search of a semiclassical sector of the theory (that would relate the theory with smooth solutions to the Einstein's equation) remain as the main open problems of the theory. Besides, the identification of a cosmological sector within LQG in order to make contact with the results of LQC---and, in the semiclassical limit, with relativistic cosmology---is still missing~\cite{ashtekarShortReviewLoop2021}.

The study of simple (truncated) models within the full theory has proved to be extremely useful. The truncation to a fixed graph is described by the corresponding holonomy-flux phase space on the graph and its quantization leads to the spin networks with support on that graph. The simplest nontrivial graphs are the \mbox{two-vertex}, $N$-edge graphs (the so-called two-vertex model, for brevity)~\cite{Rovelli_Vidotto_SteppingOutHomogeneity2008, Borja_Freidel_Garay_Livine_2010_Return, Borja_Diaz-Polo_Garay_Livine_2010_Dynamics, Borja:2011ha, Aranguren_Garay_Livine_2022,Livine_Martin-Benito_ClassicalSettingEffective2013}. The two-vertex model can be interpreted as two polyhedra with identical areas but, in general, with different volume. The classical dynamics of these polyhedra has been studied analytically in a symmetry-reduced sector (by a global $\text{U}(N)$ symmetry)~\cite{Borja_Freidel_Garay_Livine_2010_Return} and, more recently, numerically in the general case~\cite{Aranguren_Garay_Livine_2022}. In both situations a cosmological behavior is observed, presenting oscillatory and divergent regimes that avoid any singularities. Besides, in~\cite{Livine_Martin-Benito_ClassicalSettingEffective2013} it was shown that, in the low-curvature regime, the $\text{U}(N)$-reduced \mbox{two-vertex} dynamics is related to the classical dynamics of a Friedmann-Lemaître-Robertson-Walker (FLRW) universe   with a cosmological constant.

We will describe the two-vertex model using the spinorial formalism for LQG~\cite{Livine2011, Borja_Freidel_Garay_Livine_2010_Return, Livine:2013zha,Dupuis:2011dyz}, a description of the phase space of each edge using two spinors, one for each vertex of the edge.  These spinors satisfy two constraints: the matching constraint---which ensures that there is only one SU(2) irreducible representation (one spin) on each edge---and the closure constraint---which ensures the SU(2) invariance of the theory. 
The  spinorial formalism allows for a simple description of the model in terms of $2N$ spinors that, upon quantization, recovers the Hilbert space of LQG on the graph~\cite{Livine2011}. Therefore, we can study the classical dynamics of the model using the spinorial formalism in order to  gain insight into full LQG, as it will give us a first contribution to the quantum dynamics of states peaked on classical trajectories.
Moreover, using Minkowski's theorem for convex polyhedra~\cite{aleksandrov_convex_polyhedra}, the spinors provide a parametrization of the framework described by the twisted geometries~\cite{freidelTwistedGeometriesGeometric2010a, freidelTwistorsTwistedGeometries2010,Dupuis:2011dyz}. In this framework, each vertex of the graph is associated with a polyhedron and the edges indicate adjacent faces with the same area but, in general, different shapes. 
In this way, the classical counterpart (parametrized by the spinors) associated with a spin network gives us an unconventional discretization of space in which the polyhedra do not fit, i.e., they are twisted with respect to each other.  

There are no closed general formulas to compute the volume of these polyhedra (apart from the tetrahedron). Therefore, we need to resort to numerical methods~\cite{bianchiPolyhedraLoopQuantum2011, sellaroliAlgorithmReconstructConvex2017} or to approximations to the volume formula. One such approximation is given in terms of the quadrupole moment for discrete surfaces, introduced in~\cite{goellerProbingShapeQuantum2018} and studied numerically in~\cite{Aranguren_Garay_Livine_2022}, whose determinant qualitatively replicates the behavior of the volume of the polyhedron.
In this paper we analytically calculate  the equation of motion for the volume approximated by the quadrupole moment.

The $\text{U}(N)$-symmetric sector of the two-vertex model contains just one degree of freedom corresponding to the twist angle and its canonical conjugate momentum, i.e., the total area of the polyhedra. Comparing the evolution of the quadrupole approximation to the volume and the total area, we find that they precisely follow the evolution one expects from a homogeneous and isotropic evolution. We also find that a generalization of the two-vertex Hamiltonian constraint in the form of the addition of a function of the area (that preserves the symmetries of the reduced model) effectively takes into account matter content and, furthermore, fully determines the equation of state.

Finally, we show that the two-vertex model is indeed an LQG truncation that effectively describes the LQC old dynamics \cite{Bojowald:1999ts,Bojowald:2001xe} with  arbitrary barotropic perfect fluids. Furthermore, a suitable modification of the Poisson bracket structure of the U($N$)-symmetric model leads to the LQC improved dynamics \cite{Ashtekar_Pawlowski_QuantumNature2006}.

The paper is structured as follows. In Sec.~\ref{sec:spinorial_twisted} we introduce the spinorial formalism and the twisted geometries. We will use both of them in Sec.~\ref{sec:two_vertex_model} to describe the \mbox{two-vertex} model and its dynamics. 
We will also use the notion of quadrupole moment of a discrete surface to study analytically the volume and its evolution. In Sec.~\ref{sec:U(N)_symmetry} we introduce the global $\text{U}(N)$ symmetry and study the dynamics on the reduced sector. In  Sec.~\ref{sec:lqcfrom2v}, we show that the two-vertex model directly provides the FRLW cosmology and the LQC effective dynamics. We finally summarize and conclude in Sec.~\ref{sec:conclusions}.

\textit{Index notation.} 
$a,b\ldots=1,2,3$ are spatial indices; \mbox{$i,j\ldots=1,2...,N$} denote the edges of the two-vertex graph; \mbox{$I,J\ldots=1,2,3$} are $\mathfrak{su}(2)$ indices; and \mbox{$A,B = 0,1$} are spinorial indices.

\section{Spinorial formalism and twisted geometries} \label{sec:spinorial_twisted}

In this section we will introduce the spinorial and the twisted geometries formalisms~\cite{Livine2011, Borja_Freidel_Garay_Livine_2010_Return, freidelTwistedGeometriesGeometric2010a, freidelTwistorsTwistedGeometries2010}, which we will use to describe our model and interpret it geometrically from a classical point of view as a discretization of space.

\subsection{Spinorial formalism for LQG} \label{subsec:spinorial}

The phase space of LQG is usually expressed in terms of holonomies and fluxes~\cite{thiemannModernCanonicalQuantum2008}. More specifically, given a fixed graph with $N$ edges, the usual formulation assigns to each edge $i$ (with $i = 1, \ldots , N$) a pair of variables ($g_i$, $\vec{X}_i$), where $g_i \in \text{SU}(2)$ is the holonomy of the Ashtekar connection along $i$ and $\vec{X}_i \cdot \vec{\sigma} \in \mathfrak{su}\mathrm{(2)}$ is the flux of the densitized triad on a surface dual to $i$, with $\vec{X}_i \in \mathbb{R}^3$, and $\vec{\sigma} = (\sigma^1, \sigma^2,\sigma^3)$ represents the three Pauli matrices, normalized to $(\sigma^I)^2=\mathbb{I}$ for each $I=1,2,3$. 
Thus, the classical phase space $\mathrm{SU(2)} \times \mathfrak{su}\mathrm{(2)}$ of each edge $i$ is isomorphic to $T^{*} \mathrm{SU(2)}$, the cotangent bundle of SU(2)~\cite{Livine2011}. Over the last decade, an alternative parametrization has been developed in terms of two spinors $\ket{z^{s}_i}, \ket{z^{t}_i} \in \mathbb{C}^2$, which we will call source  and target spinors, respectively\footnote{Note that, although we will use bra-ket notation for the spinors, the whole formalism discussed in this paper is purely classical.}. Explicitly, the components of these spinors are
\begin{equation}
    \ket{z^s_i} = \begin{pmatrix}
        z^{s0}_i\\
        z^{s1}_i
    \end{pmatrix}, \quad    
\end{equation}
and similarly for $\ket{z^{t}}$. For convenience, we define their conjugate and dual spinors~\cite{Livine2011}
\begin{equation}
        \bra{z^s_i} = \big( \Bar{z}_i^{s0}, \; \Bar{z}_i^{s1} \big), \quad
        |z_i^s] = -i \sigma^{2} \ket{\bar{z}_i^s} = \begin{pmatrix}
        - \Bar{z}_i^{s1}\\
        \Bar{z}_i^{s0}
    \end{pmatrix},
\end{equation}
and similarly for $\ket{z^{t}}$.

We endow the spinors with canonical commutation relations, defined by
\begin{equation}
\label{eq:commutation}
    \big\{ z^{s A}_i , \bar{z}^{s B}_j \big\} = \big\{ z^{t A}_i , \bar{z}^{t B}_j \big\} = -i \delta^{AB} \delta_{ij},
\end{equation}
 and the remaining Poisson brackets are zero.
This Poisson bracket algebra is invariant under the following transformation: 
\begin{equation}
|z^{s,t\, A}_i|\to |z^{s,t\, A}_i|/\sqrt\beta,\qquad \arg z^{s,t\, A}_i\to \beta\arg\,z^{s,t\, A}_i,
\label{eq:immirziz}\end{equation}
where $\beta\in \mathbb{R}^+$ is the Barbero-Immirzi parameter (this will become clear below, when looking at the holonomy-flux algebra). This ambiguity in the choice of canonical variables will be relevant in Sec.~\ref{sec:lqcfrom2v}.

This parametrization of $T^{*} \mathrm{SU(2)}$ in terms of two spinors is the classical analog to Schwinger's representation of SU(2) by two decoupled harmonic oscillators ~\cite{GirelliLivine_2005}. According to this representation, by canonically quantizing the spinors with the commutation relations~\eqref{eq:commutation}, we will obtain two harmonic oscillators on each edge, describing the SU(2) irreducible representation associated with such edge~\cite{GirelliLivine_2005,Borja_Diaz-Polo_Garay_Livine_2010_Dynamics, Livine2011}.

From these spinors we can reconstruct the holonomies $g_i$ and the fluxes $\vec{X}^{s,t}_i $~\cite{Livine2011}:
\begin{gather}
\label{eq:elementoSU2}
    g_i := \frac{|z_i^t] \bra{z_i^s} - \ket{z_i^t} [z_i^s|}{\sqrt{\braket{z_i^t|z_i^t} \braket{z_i^s|z_i^s}}},\\[4pt]
\label{eq:vector}
    \hspace*{-1mm}\Vec{X}^{s,t}_i\! :=\! \frac{1}{2}\! \bra{z_i^{s,t}} \Vec{\sigma} \ket{z_i^{s,t}}, \quad\!\! X^{s,t}_i := |\Vec{X}^{s,t}_i| \!=\! \frac{1}{2} \!\braket{z_i^{s,t}|z_i^{s,t}}.
\end{gather}
The Poisson bracket between the holonomies and the fluxes is 
\begin{equation}
\{g_i,\vec X_j^{s,t}\}= \pm ig_i\vec\sigma\delta_{ij}/2.
\end{equation}
The Barbero-Immirzi parameter $\beta$ divides the flux and nontrivially modifies the holonomy, in such a way that the Poisson bracket is invariant under the transformation~\eqref{eq:immirziz}  as happens in LQG.
The action of $g_i$ associates the (normalized) spinors on one vertex with the dual (normalized) spinors on the other vertex:
\begin{equation}
    g_i \frac{\ket{z^s_i}}{\sqrt{\braket{z^s_i|z^s_i}}} = \frac{|z^t_i]}{\sqrt{\braket{z^t_i|z^t_i}}}, \qquad 
    g_i 
     \frac{|z^s_i]}{\sqrt{\braket{z^s_i|z^s_i}}}= -
    \frac{\ket{z^t_i}}{\sqrt{\braket{z^t_i|z^t_i}}}.
\end{equation}

Each edge of a spin network   carries an irreducible representation of SU(2). This representation must be the same seen from each of the two vertices that share the edge. In the spinorial formalism, this condition translates into the fact that the norm of the spinors $\ket{z^s_i}$ and $\ket{z^t_i}$ must be the same, for which we impose the matching constraint
\begin{equation}
\label{eq:matching}
    \mathcal{C}_i = \braket{z^{s}_i|z^{s}_i} - \braket{z^{{t}}_i|z^{{t}}_i} = 2 \left(X^s_i - X^t_i \right) = 0.
\end{equation}

This constraint generates (via Poisson brackets) U(1) transformations on the spinors. This will be of particular importance for the geometrical interpretation of our graphs, as we will see in the next subsection. It can be shown that there is a diffeomorphism between the phase space of each edge (removing the zero-norm vectors) $T^{*} \mathrm{SU(2)} - \{|\vec{X}|=0\}$ on the one hand, and the space $\mathbb{C}^2 \times \mathbb{C}^2$ of the spinors at each end of the edge reduced by the matching constraint on the other hand~\cite{Livine2011}. 

The components of the vectors $\vec{X}^{s,t}_i$ form two Lie algebras $\mathfrak{su}\mathrm{(2)}$ via Poisson brackets and their action (via Poisson brackets) on the spinors $\ket{z^{s,t}_i}$ generates SU(2) transformations. Thus, the closure vectors, defined as the sum of all the flux vectors in each vertex, $\vec{\mathcal{X}}^{s,t} = \sum_{k=1}^N \Vec{X}^{s,t}_k$, generate global SU(2) transformations in all the spinors in that vertex. Since we seek to obtain an SU(2)-invariant theory, our observables must be invariant under the transformations generated by the closure vectors. The SU(2) invariance of the intertwiners in LQG is recovered upon quantization if the so-called closure constraints have been previously imposed~\cite{Livine2011} 
\begin{equation}
\label{eq:cierre}
\vec{\mathcal{X}}^{s,t} = \sum_{i=1}^N \Vec{X}^{s,t}_i = 0.
\end{equation}

We will use this formalism in the subsequent sections in order to study the dynamics of a truncated model within the LQG.

\subsection{Twisted geometries}\label{subsec:twisted}

Spin networks are eigenstates of the area and volume operators, leading to a notion of discrete geometry given by a graph with chunks of volume attached to the vertices and quanta of area associated with the edges. This notion arises after quantization, but we can look for an equivalent framework that allows us to understand geometrically the graphs at the classical level. One such framework is given by the twisted geometries~\cite{freidelTwistedGeometriesGeometric2010a,freidelTwistorsTwistedGeometries2010}.
 
\begin{figure}
\centering
    \subfloat[][]{\label{fig:aranha} \includegraphics[width=0.3\textwidth]{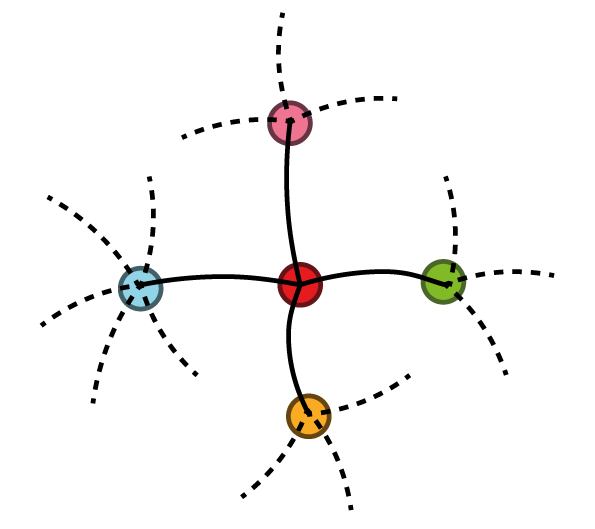}}\\
    \subfloat[][]{\label{fig:triangulitos} \includegraphics[width=0.3\textwidth]{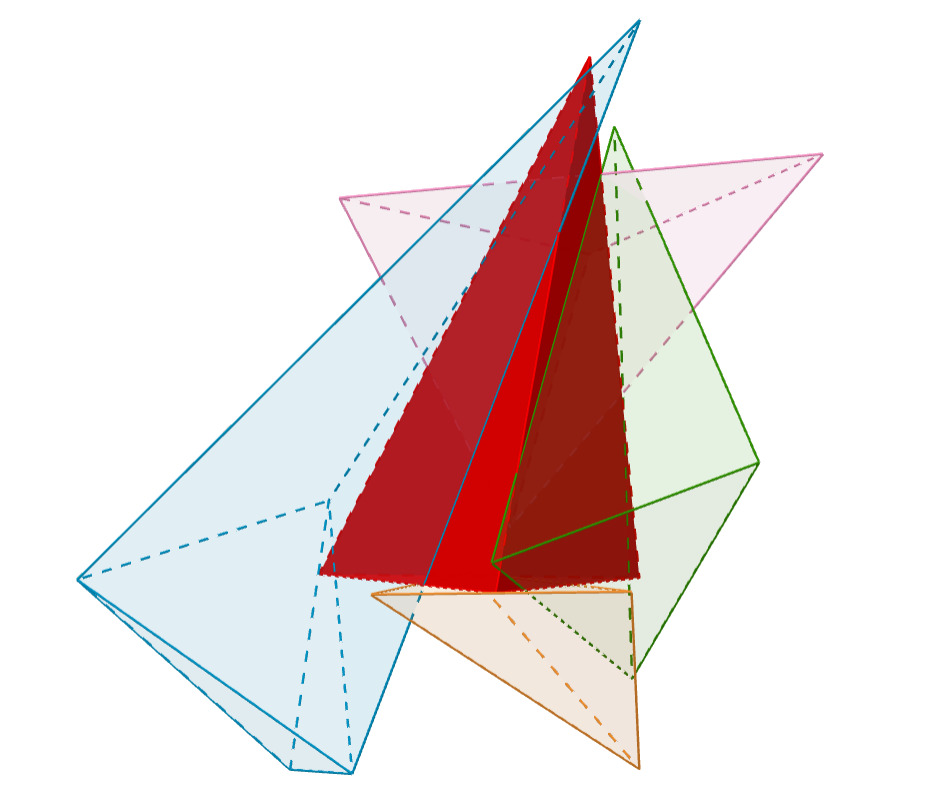}}
        \caption{(a) Example of a portion of a graph and (b) representation of the dual polyhedra to this graph given by the twisted geometries. We indicate with the same colors in both figures the edges and their corresponding polyhedra. We have taken a low number of edges for simplicity.}
        \label{fig:aranha_triangulitos}
\end{figure}

The closure constraint~\eqref{eq:cierre} on noncoplanar vectors is precisely the condition required by Minkowski's theorem~\cite{aleksandrov_convex_polyhedra}. Therefore, these vectors define a unique polyhedron at each vertex, whose faces are orthogonal to the vectors and its areas equal their norms.  
The faces associated with the same edge but different vertex are understood as adjacent, and the matching constraint~\eqref{eq:matching} causes both faces on each edge to have the same area.
However, there is no restriction on the shape of the faces. The faces associated with the same edge do not necessarily have to fit together, giving a notion of twist between adjacent faces. The twist on each edge is parametrized by the twist angle $\xi_i \in [-\pi, \pi]$. Consequently, a graph will describe a discretization of space, in which the polyhedra that share an edge will have an adjacent face with the same area but different shapes. For this reason this formalism is known as twisted geometries~\cite{freidelTwistedGeometriesGeometric2010a, freidelTwistorsTwistedGeometries2010}. 
According to this formalism, the phase space corresponding to each edge $i$ connecting two vertices is described by $(\hat{X}^s_i, \hat{X}^t_i, X_i, \xi_i)$, where $\hat{X}^s_i$ and $\hat{X}^t_i$ are the normal vectors to the source and target faces corresponding to the edge $i$; $X_i : = X_i^s = X_i^t$ is the area of the faces associated with the edge $i$; and $\xi_i$ is the twist angle of that edge. See Fig. (\ref{fig:aranha_triangulitos}) for a graphical representation of the polyhedra associated with a given graph.

Once the matching~\eqref{eq:matching} and closure~\eqref{eq:cierre} constraints are imposed, the phase space of the twisted geometries proves to be isomorphic to the gauge-invariant classical phase space of each edge. The quantization of this phase space gives us the Hilbert space of LQG~\cite{freidelTwistedGeometriesGeometric2010a, freidelTwistorsTwistedGeometries2010}.

Alternatively, one could impose additional constraints that fix the shapes of the adjacent faces, eliminating the twist. Imposing these constraints on graphs dual to triangulations, the twisted geometries   reduce  to Regge geometries~\cite{Regge_1961,freidelTwistorsTwistedGeometries2010, dittrichAreaAngleVariables2008,rovelliGeometryLoopQuantum2010}. In the past, there have been attempts to quantize Regge geometries without success~\cite{Immirzi_1996_Regge}. In this sense, the degree of freedom associated with the possibility of having faces with different shapes on the same edge allows us to quantize the twisted geometries to obtain   LQG~\cite{freidelTwistorsTwistedGeometries2010}.

\section{Two-vertex model}\label{sec:two_vertex_model}

We will now apply the formalisms explained in the previous section to describe the two-vertex model  and   introduce the dynamics proposed in~\cite{Borja_Freidel_Garay_Livine_2010_Return, Aranguren_Garay_Livine_2022}. We will then study the evolution of the volume of the dual polyhedra to the two-vertex graph. To do this, we will use the geometric quadrupole proposed in~\cite{goellerProbingShapeQuantum2018}, which allows us to obtain an analytical expression that approximates the volume of the polyhedra and study its evolution.

\subsection{Description and dynamics of the two-vertex model}\label{subsec:two_vertex_description}

\begin{figure}
\centering
    \includegraphics[width=0.4\textwidth]{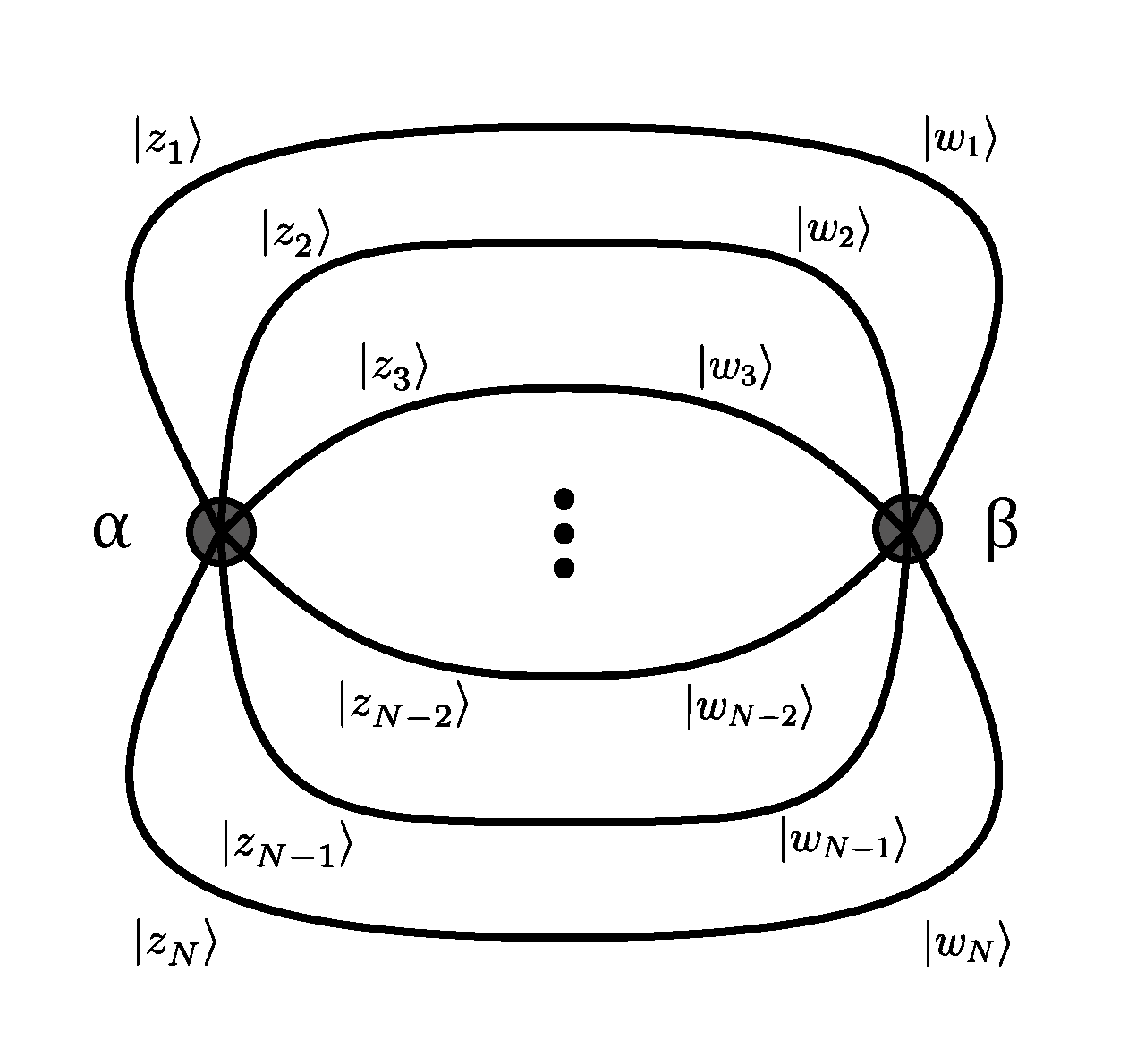}
    \caption{Two-vertex graph, labeled by $\alpha$ and $\beta$, with $N$ edges, and their respective spinors.} 
        \label{fig:dosvertices}
\end{figure}

Let us   consider the two-vertex model. We  will label the vertices\footnote{Note that the concept of source and target vertices depends on the edge, while the labels $\alpha$ and $\beta$ characterize concrete vertices. This notation is more convenient for the two-vertex model.} 
by $\alpha$ and $\beta$ 
(see Fig.~\ref{fig:dosvertices})~\cite{Aranguren_Garay_Livine_2022, Borja_Diaz-Polo_Garay_Livine_2010_Dynamics, Borja_Freidel_Garay_Livine_2010_Return}. 
Following the standard notation in the literature,  we will denote spinors at vertex $\alpha$ as $\ket{z_i} := \ket{z^{\alpha}_i}$, spinors at vertex $\beta$ as $\ket{w_i} := \ket{z^{\beta}_i}$, and the vectors at each vertex as \mbox{$ \vec{X}_i := \vec{X}_i^{\alpha}$} and $\vec{Y}_i := \vec{X}_i^{\beta}$.

The simplicity of this model has made it possible to implement a nontrivial dynamics. With this aim, we  consider the SU(2)-invariant observables~\cite{Borja_Diaz-Polo_Garay_Livine_2010_Dynamics, Borja_Freidel_Garay_Livine_2010_Return}  (at the vertex $\alpha$)
\begin{equation}
\label{eq:operadores_invSU(2)}
    E_{kl}^{\alpha} = \braket{z_k|z_l}, \quad F_{kl}^{\alpha} = [z_k|z_l\ra, \quad \bar{F}_{kl}^{\alpha} = \la z_l|z_k],
\end{equation}
and likewise for the vertex $\beta$.
Using these observables, we   construct the Hamiltonian $H = \mathcal{M} C$ proposed in~\cite{Borja_Diaz-Polo_Garay_Livine_2010_Dynamics, Borja_Freidel_Garay_Livine_2010_Return}, where
\begin{equation}
    \label{eq:hamiltoniano}
    C = \sum_{k,l=1}^N \big[\lambda E_{kl}^{\alpha} E_{kl}^{\beta} + \text{Re}(\gamma F_{kl}^{\alpha} F_{kl}^{\beta} )
    \big],
\end{equation}
$\lambda \in \mathbb{R}, \gamma \in \mathbb{C}$ are constants with units of inverse length\footnote{\label{fn:volume} Note   that $C$ is a Hamiltonian constraint that would come from integrating the Hamiltonian density from LQG  over a fiducial spatial volume and therefore the coupling constants carry this information.}, and the lapse $\mathcal{M}$ is a Lagrange multiplier that enforces the Hamiltonian constraint $C = 0$. This Hamiltonian is constructed to the lowest order in the operators~\eqref{eq:operadores_invSU(2)} (treating both vertices equally). It is SU(2)-invariant, i.e., $\{H, \vec{\mathcal{X}}\} = \{H, \vec{\mathcal{Y}}\} = 0$, where $\vec{\mathcal{X}}$ and $\vec{\mathcal{Y}}$ are the closure vector~\eqref{eq:cierre} at the vertices $\alpha$ and $\beta$, respectively. Furthermore, it satisfies the matching constraint, so it is U(1) invariant on every edge, i.e. $\{H,\mathcal{C}_i\} = 0$. 
For simplicity, the coupling constants $\lambda$ and $\gamma$ are taken edge independent but, in principle, they could also be different on each edge~\cite{Borja_Freidel_Garay_Livine_2010_Return, Aranguren_Garay_Livine_2022}. Nevertheless, the symmetry reduction to the $\text{U}(N)$-invariant sector will impose edge independence on the constants~\cite{Borja_Freidel_Garay_Livine_2010_Return}, as it is already assumed in~\eqref{eq:hamiltoniano}.

\subsection{Volume and quadrupole of a surface}\label{subsec:quadrupole_volume}

In Sec.~\ref{subsec:twisted} it was shown that, as a consequence of the closure constraint, the interpretation of twisted geometries assigns a dual polyhedron to each of the two  vertices of our graph, thus providing a classical notion of areas at the edges and volumes at the vertices. In this section we will focus on the polyhedron associated with the vertex~$\alpha$. The reasoning is completely analogous for the vertex $\beta$.

Minkowski's theorem~\cite{aleksandrov_convex_polyhedra} ensures the existence of a polyhedron associated with a set of vectors satisfying the closure constraint, but it does not give a prescription for the construction of the polyhedron. Although the polyhedron-reconstruction algorithm has recently been presented in~\cite{bianchiPolyhedraLoopQuantum2011,sellaroliAlgorithmReconstructConvex2017} and used in~\cite{Aranguren_Garay_Livine_2022} in order to compute the volume numerically, there is no analytical formula for the volume of a polyhedron with more than four faces in terms of the normal vectors to its faces. The tetrahedron ($N = 4$) is an exception, and the volume squared is given by~\cite{goellerProbingShapeQuantum2018}:
\begin{equation}
    V^2 = \frac{2}{9} \left| \vec{X}_1 \cdot \left(\vec{X}_2 \wedge \vec{X}_3 \right) \right|.
    \label{eq:volumen_tetraedro}
\end{equation}

In order to analytically study  the behavior of the volume of general polyhedra (with $N>4$), we can use a geometric multipole expansion for closed surfaces~\cite{goellerProbingShapeQuantum2018}. This expansion defines the monopole as the surface area and the dipole as the center of mass of the surface (which becomes null when the closure constraint~\eqref{eq:cierre} is applied). Additionally, the geometric quadrupole of the polyhedron~$\alpha$ is defined as~\cite{goellerProbingShapeQuantum2018}
\begin{equation}
    \label{eq:cuadrupolo}
    T^{IJ} = \sum_{k=1}^N \frac{X_k^I X_k^J}{X_k} .
\end{equation}
This quadrupole gives us basic information about the shape of the surface: its eigenvalues have a nontrivial relation with the principal radii of the ellipsoid that best approximates the surface in question, i.e. the ellipsoid with the same quadrupole~\cite{goellerProbingShapeQuantum2018}.

Given that $X_k$ represents the area of the $k$th face, the determinant of the geometrical quadrupole has an interpretation as a volume squared. Previous papers have explored the use of this determinant as an alternative to the   volume squared of the polyhedron~\cite{goellerProbingShapeQuantum2018,Aranguren_Garay_Livine_2022}, showing that both the determinant of the quadrupole and the volume squared (numerically calculated using the polyhedron-reconstruction algorithm presented in~\cite{bianchiPolyhedraLoopQuantum2011,sellaroliAlgorithmReconstructConvex2017}), although they do not coincide in value, have the same behavior in terms of their growth and extrema~\cite{Aranguren_Garay_Livine_2022}. 
Therefore, the determinant of the quadrupole may be a good tool to study the evolution of the volume, for which we define the following function:
\begin{equation}
\label{eq:volumen_aprox}
    \tilde{V}^2 = \frac{4\pi}{3} \det T = \frac{2\pi}{9}\varepsilon^{IJK} \varepsilon^{LMN} T_{IL}T_{JM}T_{KN},
\end{equation}
where $\varepsilon_{IJK}$ is the Levi-Civita symbol (with $\varepsilon_{123} = 1$).
We will then use the function~\eqref{eq:volumen_aprox}, which we will refer to as the approximate volume squared, to approximately study the evolution of our polyhedra.

\subsection{Evolution of the volume} \label{subsec:two_vertex_volume}

The evolution of the volume under the dynamics given by the Hamiltonian~\eqref{eq:hamiltoniano} has been numerically studied in~\cite{Aranguren_Garay_Livine_2022} using the polyhedron-reconstruction algorithm~\cite{bianchiPolyhedraLoopQuantum2011,sellaroliAlgorithmReconstructConvex2017}. In order to obtain an analytical expression for the volume and its evolution, we will use the volume 
approximation~\eqref{eq:volumen_aprox}. Indeed, after carrying out a straightforward but lengthy calculation, we get the evolution of the volume of the polyhedron associated with the \mbox{vertex $\alpha$}:
\begin{align}
\label{eq:evol_volumen}
     &\dot{\tilde{V}} :=  \{H,\tilde{V}\} =  \frac{\pi}{3\tilde{V}} \varepsilon^{IJK} \varepsilon^{LMN} \dot{T}_{IL}T_{JM}T_{KN} \\
     &= \frac{\pi \mathcal{M}}{3\tilde{V}} \mathrm{Im} \!\!\!\! \sum_{i,j,k,l=1}^N \!\!\!\!\!\frac{\left(\vec{X}_i \!\wedge\! \vec{X}_j\right)\! \cdot\! \vec{X}_l}{X_i X_j X_l}
      \Bigg\{\! \frac{\left(\vec{X}_i \!\wedge\! \vec{X}_j\right)\! \cdot \!\vec{X}_l}{X_l} \!\left(\! \lambda E_{kl}^{\alpha} E_{kl}^{\beta} \right. \nonumber \\
     & \left. +  \gamma F_{kl}^{\alpha}F_{kl}^{\beta}\right) 
    \!- 2 \left( \vec{X}_i \!\wedge \!\vec{X}_j\right)_I \! \left(\lambda G_{kl}^I E_{kl}^{\beta} +   \gamma S_{kl}^I F_{kl}^{\beta} \right) \Bigg\},\nonumber
\end{align}
where the dot denotes derivative with respect to the evolution parameter $t$ associated with the Hamiltonian~$H$, and we have introduced the following (non-gauge-invariant) functions:
    \begin{equation}
    S^I_{kl}=[z_k|\sigma^I|z_l\ra=S^I_{lk},\quad 
    G^I_{kl}=\la z_k|\sigma^I|z_l\ra=\bar{G}^I_{lk}.
    \end{equation}
    
We observe in the analytical expression for the evolution of the volume~\eqref{eq:evol_volumen} that a degenerate polyhedron (i.e. with all its faces parallel to each other) has null volume throughout the entire evolution. Similarly, we can obtain the evolution of the polyhedron associated with the vertex $\beta$ by substituting in Eq.~\eqref{eq:evol_volumen} the functions and vectors defined on $\alpha$ by the corresponding ones defined on $\beta$ and vice versa. 

The behavior of both the evolution of the approximate volume~\eqref{eq:volumen_aprox} and the exact one (calculated numerically) is qualitatively similar, as shown in ~\cite{Aranguren_Garay_Livine_2022}. Note that, even though the volume of the two polyhedra will differ in general, the matching constraint ensures that the area of both polyhedra remains equal to each other.

\section{U($N$)-invariant sector of the two-vertex model}\label{sec:U(N)_symmetry}

The two-vertex model presents an invariant sector under a global U$(N)$ symmetry (whose generators involve operators acting on both vertices)~\cite{Borja_Diaz-Polo_Garay_Livine_2010_Dynamics, Borja_Freidel_Garay_Livine_2010_Return}. One of the main results in this sector is that the equations of motion of the total area of the polyhedra are mathematically analogous to the equations of motion of volume in LQC~\cite{Borja_Diaz-Polo_Garay_Livine_2010_Dynamics}. This has lead to an interpretation of the two-vertex model as a cosmological model. In this section we will review the U($N$) symmetry and we will study the dynamics of the two-vertex model (fully described with spinors) on the U($N$)-invariant sector.

\subsection{U($N$) symmetry} \label{subsec:U(N)_symmetry}

The $\mathrm{U(1)}^N$ symmetry generated by the action of the matching constraints $\mathcal{C}_i$ on each edge is part of a greater U($N$) symmetry generated via Poisson brackets by~\cite{Borja_Diaz-Polo_Garay_Livine_2010_Dynamics}
\begin{equation}
   \mathcal{E}_{ij} = E_{ij}^{\alpha} - E_{ji}^{\beta}.
\end{equation} 
Indeed, we observe that the matching constraint associated with the edge $k$ is simply the $k$th element $\mathcal{C}_k = \mathcal{E}_{kk}$ of the diagonal of $\mathcal{E}_{ij}$. 

The generators $\mathcal{E}_{ij}$ are constants of motion, given that $\{\mathcal{E}_{ij}, H\} = 0$. Conversely, the Hamiltonian of the system is invariant under the U($N$) transformation generated by~$\mathcal{E}_{ij}$. These transformations will be symmetries of the system only if the constraints (in this case the matching and closure constraints) are invariant under the action of $\mathcal{E}_{ij}$. If not, applying the U($N$) transformation generated by $\mathcal{E}_{ij}$ on configurations satisfying the constraints would lead to configurations not satisfying them. 
A simple calculation shows that, while the closure constraint is invariant, $\{\mathcal{E}_{ij}, \vec{\mathcal{X}}\} = 0$, the action of $\mathcal{E}_{ij}$ on the matching constraint~\eqref{eq:matching} is nontrivial,
\begin{equation}
    \left\{\mathcal{C}_k,\mathcal{E}_{ij}\right\} = i\left[(E_{ik}^{\alpha} - E_{ki}^{\beta})\delta_{kj}-(E_{kj}^{\alpha} - E_{jk}^{\beta})\delta_{ki}\right].\\
\end{equation}
That is, the action of $\mathcal{E}_{ij}$ on a pair of spinors that initially satisfy the matching constraint~\eqref{eq:matching} will generally yield another pair of spinors that do not. However, we can work on the subspace of spinors that satisfy \mbox{$\left\{\mathcal{C}_k, \mathcal{E}_{ij} \right\} = 0$}, restricting our phase space to those configurations that satisfy
\begin{equation}
    (E_{ik}^{\alpha} - E_{ki}^{\beta})\delta_{kj} = (E_{kj}^{\alpha} - E_{jk}^{\beta})\delta_{ki}.
\end{equation}
The diagonal case $i=j$ is trivial (the matching constraint~$\mathcal{C}_k$ commutes with itself, as it generates uncoupled U(1) transformations). The case $i \neq j = k$ gives us the constraint
\begin{equation}
\label{eq:Eija=Ejib}
    \mathcal{E}_{ij} =  E_{ij}^{\alpha} - E_{ji}^{\beta} = 0,
\end{equation}
which implies that
\begin{equation}
    \la z_i|z_j\ra = \la \bar{z}_j|\bar{z}_i\ra = \la w_j|w_i\ra.
\end{equation}

To satisfy this constraint, the spinors must be related by a unitary matrix $U \in \mathrm{U(2)}$ such that $\ket{w_i} = U \ket{\bar{z}_i}$. We can write any element   $U\in \text{U}(2)$ as $U = e^{-i\varphi/2} h$, where $h \in \mathrm{SU(2)}$ and $\varphi \in [-\pi, \pi]$ is an arbitrary phase. The theory is SU(2) invariant, so the SU(2) part of $U$ is pure gauge and we can fix it without loss of generality. Doing so, the only part that remains is the phase $e^{-i\varphi/2} \in \mathrm{U(1)}$. Therefore, we find that the constraint $\mathcal{E}_{ij} = 0$ imposes the following relation between the spinors on each edge:
\begin{equation} \label{eq:espinores_fase_U(N)}
    \ket{w_{i}} = e^{-i\varphi/2} \ket{\bar{z}_{i}}.
\end{equation}

The Hamiltonian is invariant under the action of $\mathcal{E}_{ij}$. Thus initial conditions satisfying the U($N$) symmetry (that is, pairs of initial spinors satisfying ~\eqref{eq:espinores_fase_U(N)}) will ensure U($N$) symmetry throughout the whole evolution. Expressing the action on the phase space surface given by $\mathcal{E}_{ij}=0$ and applying the relation~\eqref{eq:espinores_fase_U(N)}, we find that only one degree of freedom remains. If we choose that degree of freedom to be the angle $\varphi $, the canonical conjugate variable will be the total area of the polyhedra~\cite{Borja_Freidel_Garay_Livine_2010_Return} \mbox{$\area = \sum_{i=1}^N X_i = \sum_{i=1}^N Y_i$}, i.e. 
\begin{equation}
\{\varphi,\area\}=1.
\end{equation}

At this point, it is important to note that the ambiguity \eqref{eq:immirziz} parametrized by the Barbero-Immirzi parameter~$\beta$ translates directly to these variables  in the form of the invariance of the Poisson brackets between them under constant Weyl transformations with parameter $\beta$, i.e. $\area\to \area/\beta$, $\varphi\to \beta\varphi$.

Then the Hamiltonian constraint~\eqref{eq:hamiltoniano} on the surface $\mathcal{E}_{ij}=0$ in terms of the canonical pair $(\varphi,\area)$ reduces to
\begin{equation}
\label{eq:hamiltoniano_U(N)}
   \Cred = 2  \area^2 (\lambda +  \gamma \cos\varphi),
\end{equation}
where the symbol $\breve{}$ denotes U($N$)-symmetry reduction. We can choose $\gamma \in \mathbb{R}^{+}$ without loss of generality, as any phase can be absorbed in the definition of $\varphi$. After imposing the symmetry and the matching constraint, our Hamiltonian~\eqref{eq:hamiltoniano_U(N)} only depends on two variables: the   total area $\area$ of the polyhedron and the angle $\varphi$. We observe that both variables are global, in the sense that they have no information about different edges and vertices. Because of that, the relation~\eqref{eq:espinores_fase_U(N)} has been interpreted as a natural way of imposing homogeneity and isotropy on the graph~\cite{Borja_Diaz-Polo_Garay_Livine_2010_Dynamics, Borja_Freidel_Garay_Livine_2010_Return}.

Imposing the constraint $\Cred = 0$ means that either $\cos \varphi = -\lambda/ \gamma$ or $\area = 0$. The latter can always be imposed, while the former is only possible if $|\lambda|<  \gamma$. As we commented in Sec.~\ref{subsec:two_vertex_volume}, the null-volume (and thus the null-area) polyhedra are decoupled from the rest of the configurations. So we can focus on the case of constant~$\varphi$. Using the U($N$) symmetry-reduced Hamiltonian constraint~\eqref{eq:hamiltoniano_U(N)}, we can derive the evolution equations for the conjugate variables $\varphi$ and $\area$ as follows:
\begin{alignat}{2}
\label{eq:evolucion_phi}
    \dot{\varphi} &= \{ \varphi, \Hred\} &&= 4\mathcal{M}  \area  [\lambda + \gamma \cos \varphi]  = 0, \\
\label{eq:evolucion_E}
    \dot{\area} &= \{ \area , \Hred\} &&=   2\mathcal{M}  \gamma\area^2  \sin  \varphi = 2\mathcal{M}  \area^2 \sqrt{ \gamma^2 - \lambda^2}.
\end{alignat}
Fixing the Lagrange multiplier $\mathcal{M} = 1$, the solution to Eq.~\eqref{eq:evolucion_E} is
\begin{equation}
\label{eq:E_t_solved}
    \area(t) = \frac{1}{\area_0^{-1} - 2 t \sqrt{ \gamma^2 - \lambda^2}},
\end{equation}
where $\area_0 = \area(t=0)$. For $|\lambda| >   \gamma $, the only possible solution to the constraint is $\area(t) = 0$ with arbitrary  constant~$\varphi$.

\subsection{Evolution of the volume}\label{sec:evol_volumen_U(N)}

After restricting our system to the U($N$)-invariant phase space sector by imposing the relation~\eqref{eq:espinores_fase_U(N)}, we can study the evolution in this sector of observables defined in the complete phase space. In this section we will study the volume, for which we have an approximation $\tilde{V}$ for a generic polyhedron with $N$ faces given by~\eqref{eq:volumen_aprox}.

After performing some algebra we find that the evolution of the approximation for the volume~\eqref{eq:evol_volumen} in the U($N$)-invariant sector is
\begin{equation}
    \dot{\tilde{V}} = 3 \mathcal{M} \area \gamma \tilde{V}\sin \varphi.
    \label{eq:evol_volumen2_U(N)}
\end{equation}

As previously mentioned, while there is no general equation to express the volume of an irregular polyhedron with $N$ faces as a function of the normal vectors to the faces, for the specific case of the tetrahedron ($N=4$) we can express the volume as a function of $\vec{X}_i$ using~\eqref{eq:volumen_tetraedro}. 
Therefore, after computing the evolution of $\vec{X}$, given by
\begin{equation} 
\dot{X}^I_i = \{X^I_i,H\} = - \, \mathcal{M}\,  \text{Im}\bigg(\sum_{k=1}^N \lambda G^I_{ki}E^\beta_{ki}+  \gamma S^I_{ki}F^\beta_{ki}\bigg),
\label{ec_evolX}
\end{equation}
it is straightforward to compute the evolution of the volume of the tetrahedron within the U($N$)-invariant sector
\begin{equation}
    \dot{V}= 3 \mathcal{M}  \area \gamma V\sin \varphi.
    \label{eq:evol_tetra_U(N)}
\end{equation}

We observe that the approximate volume $\tilde{V}$ (for any number of faces) and the exact volume $V$ for the tetrahedron have the same evolution in the reduced sector and, hence, we conclude that for the tetrahedron $\tilde{V} \propto V$. Although the approximation $\tilde{V}$ for an arbitrary number of faces $N$ still requires further investigation, the fact that its evolution on the U($N$)-invariant sector is identical to that of $V$ for the tetrahedron strengthens the argument for using $\tilde{V}$ in the general case.

On the other hand, by examining~\eqref{eq:evol_volumen2_U(N)} and~\eqref{eq:evolucion_E}, we immediately observe that, within the U($N$)-invariant sector, the approximate volume $\tilde{V}$ and $\area$ (that is the total area of the polyhedron) satisfy the relation $2 {\dot{\tilde{V}}}/{\tilde{V}} = 3 {\dot{\area}}/{\area}$, which implies:
\begin{equation}
\label{eq:area_volumen}
\tilde{V} = a_0 \area^{3/2},
\end{equation}
where $a_0$ is a constant determined by the initial conditions. Equation~\eqref{eq:area_volumen} corresponds to what one would obtain when uniformly scaling all dimensions of a three-dimensional body: when we scale all three dimensions by a factor $l$, the volume and area differentials scale as $l^3$ and $l^2$, respectively. Consequently, the relation~\eqref{eq:area_volumen} is that of a polyhedron undergoing a homogeneous and isotropic expansion, further strengthening the interpretation of this U($N$)-invariant sector as a cosmological (homogeneous and isotropic) sector. 

\subsection{Geometric interpretation}\label{sec:Interpretacion_phi}

Let us now interpret the two conjugate variables of the U($N$)-symmetric Hamiltonian~\eqref{eq:hamiltoniano_U(N)}. As shown before, the variable $\area$ is the sum of the norms of all the vectors~$\vec{X}_i$, which directly corresponds to the total area of the polyhedron in the dual graph formalism within the framework of twisted geometries. 
On the other hand, the variable~$\varphi$ defines the SU(2) holonomy~\cite{Borja_Freidel_Garay_Livine_2010_Return}. We will now provide a simple relation between $\varphi$ and the twist angles~$\xi_i$ (described in Sec.~\ref{subsec:twisted}).

The framework of twisted geometries defines a twist angle $\xi_i$ for each edge. Two possible definitions of $\xi_i$ exist, one in terms of the components $(z^0_i, w^0_i)$ of the twistor, and the other in terms of $(z^1_i, w^1_i)$~\cite{freidelTwistorsTwistedGeometries2010}:
\begin{equation}
    \xi^A_i = i \ln \frac{z^A_i w^A_i}{\bar{z}^A_i \bar{w}^A_i},
\label{eq:angulo_twist}
\end{equation}
with $A = 0,1$. Both definitions are related by a canonical transformation, making them equivalent descriptions of the twist angle~\cite{freidelTwistorsTwistedGeometries2010}. At first sight, this angle is not related to  the variable $\varphi$ defined in the U($N$)-symmetry-reduced sector. 
However, using the definition~\eqref{eq:angulo_twist} and the condition~\eqref{eq:espinores_fase_U(N)}, we find that in this sector:
\begin{equation}
    \xi_i^A = i \ln \frac{(e^{-i\varphi/2} \bar{w}_i^A) w_i^A}{(e^{i \varphi/2}w_i^A) \bar{w}_i^A} = \varphi  :=\xi.
\end{equation}
This shows that the twist angle is the same for all the edges and that $\xi^0_i = \xi^1_j $, which means that the two definitions given by~\eqref{eq:angulo_twist} coincide in the U($N$)-reduced  sector. Therefore, we can conclude that, in this case, the variable~$\varphi$ is the twist angle, which is independent of the edge.
 
In summary, we have found that the reduced sector (under the global U($N$) symmetry) of the two-vertex model depends solely on two canonically conjugate variables which  are homogeneous and isotropic, i.e. independent of the edges and vertices.
The first one is the total area $\area$ of the polyhedron, directly related to its volume through~\eqref{eq:area_volumen}. Therefore, its evolution can be interpreted as expansion. The second phase space variable is the twist angle $\varphi =  \xi$ which, as briefly discussed in the next paragraph, is related to the extrinsic curvature.
Thus, our model possesses a unique curvature that is independent of edges and a single polyhedron-area independent of vertices. Considering the results and conclusions of previous works~\cite{Borja_Freidel_Garay_Livine_2010_Return, Borja_Diaz-Polo_Garay_Livine_2010_Dynamics}, the fact that our dynamical variables admit an interpretation of area (one-to-one related with the volume by Eq.~\eqref{eq:area_volumen}) and extrinsic curvature strengthens the cosmological interpretation of this model. 

According to the proposal of Freidel and Speziale~\cite{freidelTwistorsTwistedGeometries2010}, the discrete twist angle $\xi_i$ associated with the edge $i$ can be related to the continuous extrinsic curvature by $\xi_i^2=-\ell^2 \det(\hat \ell_i^a K_a^I\sigma_I) $, where  $\ell$ is a small length scale, $\hat \ell_i^a$ is a unitary vector along the edge $i$, and $K_a$ is the $\mathfrak{su}(2)$-valued extrinsic-curvature one-form. In the U$(N)$-symmetric two-vertex model, the twist angle does not depend on the edge  $i$ as we have seen and hence the  edge information disappears from the relation between twist angle and the extrinsic curvature proposed in~\cite{freidelTwistorsTwistedGeometries2010} pointing towards an homogeneity and isotropy, which we pin down in the following section.

\section{LQC from the U$(N)$-invariant two-vertex model}\label{sec:lqcfrom2v}

In this section we show that the U$(N)$-invariant \mbox{two-vertex} model is indeed a reduction of the full LQG theory which provides LQC effective dynamics.

\subsection{The constraint equation}

Let us start by noting that the two-vertex Hamiltonian~\eqref{eq:hamiltoniano} can be modified as long as we respect the SU(2) symmetry and   preserve the commutation with the matching constraint. Thus, we   consider the following generalization of the U($N$)-reduced Hamiltonian constraint
\begin{equation}
\label{eq:hamiltonian_UN_generalisation}
    \Hredg = 2 \mathcal M  \area^2 \big[ \lambda  + \gamma \cos  \varphi  + g(\area) \big],
\end{equation}
where $g(\area)$ is an arbitrary smooth function with dimensions of inverse length. Notice that this generalized reduced Hamiltonian constraint can be obtained by imposing the $\text{U}(N)$ symmetry on different generalizations of Hamiltonian constraints of the kind of~\eqref{eq:hamiltoniano}, that is, different classical Hamiltonians for the general case may lead to the same homogeneous and isotropic reduced sector.

In order to explicitly connect with the language of cosmology and motivated by the idea that the variable $\area$ corresponds to an area, we perform the following canonical transformation
\begin{equation}
   a =  \sqrt{\area}, \qquad                       \pi_a = - 2 \sqrt{\area} \varphi.       \label{eq:canonical_transformation}
\end{equation}
The Barbero-Immirzi ambiguity~\eqref{eq:immirziz} becomes, in this case, constant Weyl invariance parametrized by $\sqrt\beta$, i.e. under the transformation $a\to a/\sqrt\beta$, $\pi_a\to\sqrt\beta\pi_a$. 
From now on, we will keep the Barbero-Immirzi parameter explicit since it will be useful in the following. Then the Hamiltonian constraint~\eqref{eq:hamiltonian_UN_generalisation} takes the form $\Hredg = \mathcal{N}  \Credg$, where
 \begin{equation}
  \Credg =  - \frac{4\gamma}{ \beta^2} a  \sin^2     \frac{\beta\pi_a}{4a}  - \frac\kappa \gamma a  +  a^3 f(a)  , \label{eq:ham_red_flrwp}
  \end{equation}
with $\kappa=-  2\gamma (\lambda+ \gamma)/ \beta^2 $, we have conveniently redefined the Lagrange multiplier as $\mathcal M= \mathcal N /  a^3 $, and $f(a)=  2 ( \beta a )^{-2}g(a^2/\beta) $ has dimensions of inverse length.
This Hamiltonian constraint~\eqref{eq:ham_red_flrwp} can be written in terms of $(a, \dot{a})$ by using the equation of motion for $a$,
  \begin{equation}
      \dot{a} = \{a,\Hredg\} =  -\mathcal{N}\frac{\gamma}{ \beta}  \sin \frac{\beta\pi_a}{2a},\label{eq:adotpia}
  \end{equation}
  and replacing it back into~\eqref{eq:ham_red_flrwp}.
An straightforward manipulation of the resulting expression allows us to write the constraint equation $\Credg =0$ as
\begin{equation}
\label{eq:friedmann_2vertex_R}
    \frac{1}{\mathcal{N}^2}\frac{\dot{a}^2}{a^2} =  \gamma\mathcal{R}  \bigg( 1 -  \frac{a^2 \beta^2}{4 \gamma} \mathcal{R}  \bigg),
\end{equation}
where 
\begin{equation}
\label{eq:R_friedmann}
    \mathcal{R}  =     f  -\frac\kappa{ \gamma a^2} .
\end{equation} 

The Hamiltonian constraint~\eqref{eq:ham_red_flrwp} and its configuration-space counterpart~\eqref{eq:friedmann_2vertex_R} will be our starting point for establishing the connection with LQC.

\subsection{FLRW cosmology as the small-twist regime of the two-vertex model} 

The small-twist limit $|\varphi|\ll 1$ is equivalent to \mbox{$|\pi_a/a|\ll 1$}. In the lowest order the Hamiltonian constraint~\eqref{eq:ham_red_flrwp} acquires the form 
\begin{equation}
   \Credg =   -\frac{\gamma}{4 } \frac{\pi_a^2} a   - \frac\kappa\gamma a +    a^3f  .
  \end{equation}
Note that in this limit the Barbero-Immirzi parameter disappears as expected.
From this Hamiltonian (or equivalently from~\eqref{eq:adotpia}) we obtain  \mbox{$\dot a=-\mathcal{N} \gamma   {\pi_a}/(2  a)$}, which, when replaced back into the constraint, yields
  \begin{equation}
 \frac{1}{\mathcal{N}^2}\frac{\dot{a}^2}{a^2} =   \gamma\mathcal{R} .
  \label{eq:fried2v}
  \end{equation}
  
This equation is precisely Friedmann's equation for a homogeneous and isotropic spacetime with spatial sections of curvature $\kappa$ and matter content given by the function $f$. Indeed, a homogeneous and isotropic universe is described by the FLRW metric
\begin{equation}
\label{eq:metrica_FLRW}
    \diff s^2 = -\mathcal{N}^2(t) \diff t^2 + a^2(t) \diff \Sigma_k^2,
\end{equation} 
where $\mathcal{N}(t)$ is the lapse function, $a(t)$ is the scale factor, $\diff\Sigma_k^2=
 {\diff r^2}/({1-k r^2}) + r^2 \diff \Omega_2^2$, $k$ is the constant that characterizes the curvature of the spatial slices, and $\diff\Omega_2^2$ is the metric of the unit two-sphere. The general-relativistic dynamics for this FLRW universe filled with a perfect fluid with density $\rho$ and pressure $p$ is described by the  Friedmann's equation 
 \begin{equation}
    \label{eq:friedmann_barotropic}
    \frac{1}{\mathcal{N}^2} \frac{\dot{a}^2}{a^2}  = \frac{8 \pi G}{3} \rho-\frac{k}{a^2},
\end{equation} 
together with the continuity equation 
\begin{equation}
    a\dot \rho=-3\dot a(\rho+p).
\end{equation}
For barotropic fluids the equation of state is such that the pressure depends only on the density and the continuity equation implies that the density is a given function of the scale factor $\rho(a)$.

In order to compare both dynamical equations~\eqref{eq:fried2v} and~\eqref{eq:friedmann_barotropic} on equal footing it is necessary to introduce a fiducial volume $\mathcal V_\Sigma$ with respect to the line element $\diff \Sigma_k^2$ (see footnote~\ref{fn:volume}). This factor naturally appears multiplying both $\rho$ and $1/G$ when considering the Hamiltonian for FLRW dynamics. In this comparison we then see that the small-twist regime of the U$(N)$-invariant \mbox{two-vertex} model describes a FLRW cosmology. The cosmological parameters and energy density are determined by the parameters $\gamma$, $\kappa$, and the function $f$ that define the two-vertex model: The gravitational constant is given by $G=3\gamma \mathcal V_\Sigma/(8\pi)$, the spatial curvature by $k= \kappa$, and the matter content is  a barotropic fluid with density $\rho= f/\mathcal V_\Sigma$ and pressure determined by the continuity equation $p=-(a^3f)'/(3\mathcal V_\Sigma a^2)$, the prime denoting derivative with respect to $a$.
 
As a simple example, the two-vertex model with 
\begin{equation}
f(a)=\sum_n f_n a^{-\alpha_n}
\end{equation}
describes a cosmology filled with a number of perfect fluids with equations of state 
\begin{equation}
p_n=w_n\rho_n, \qquad w_n= \alpha_n/3-1.
\end{equation}

As we saw in Sec.~\ref{sec:Interpretacion_phi}, $\varphi = \xi$ can be interpreted as extrinsic curvature given by 
\begin{equation}
\xi_i^2=-\ell^2 \det(\hat \ell_i^a K_a^I\sigma_I)= \ell^2\hat\ell_i^a\hat\ell_i^b K_{ac}K^c_b.
\end{equation}
The extrinsic curvature of the spatial slices of a FLRW universe is proportional to the spatial metric $q_{ab}$, i.e. $K_{ab} =   q_{ab} \dot{a}/( \mathcal{N}a)$.
This implies that $\xi_i =\ell  \dot a /(a\mathcal N)$, i.e. that it is independent of the edge~$i$ in agreement with the U$(N)$-invariance, which can be written as
\begin{equation}
\label{eq:phi_ele}
    \varphi = \frac{\dot{a} \ell}{\mathcal{N} a}.
\end{equation} 
Therefore, Freidel and Speziale's  interpretation of the twist angle in terms of the extrinsic curvature~\cite{freidelTwistorsTwistedGeometries2010} finds here an explicit realization which allows in addition to interpret $\gamma$ (and hence $G/(\beta^2\mathcal V_\Sigma)$) as the inverse of the fundamental length scale in the two-vertex model 
\begin{equation}
\ell= \beta a/ \gamma ,
\end{equation}
as can be directly seen by comparing Eq.~\eqref{eq:phi_ele} and \mbox{$\varphi=-\beta\pi_a/(2a)= \beta\dot a/( \mathcal N  \gamma)$}.

\subsection{LQC effective dynamics}\label{sec:LQC_effective_dynamics}

In the previous section we have shown that the two-vertex  dynamics describes an FLRW spacetime filled with perfect barotropic fluids in the lowest order in the twist angle, that is, in the low-curvature regime. This has allowed us to write the gravitational constant, the spatial curvature, and the matter  density in terms of the parameters $\gamma$ and $\kappa$ and the function $f$ that characterize the two-vertex model.

In the old dynamics of LQC for a universe with a free scalar field and vanishing spatial curvature, the semiclassical states (i.e. Gaussian coherent states sharply peaked at the classical FLRW trajectories in the low-curvature regime) follow the effective Friedmann's equation, as can be derived from~\cite{Ashtekar_Pawlowski_OldDynamics2006},
\begin{equation}
\label{eq:effective_friedmann_LQC}
    \frac{1}{\mathcal{N}^2} \frac{\dot{a}^2}{a^2} = \frac{8 \pi G}{3} \rho \Big(1 - \frac{a^2\rho}{\rho_\star} \Big) ,
\end{equation}
to lowest order in the Gaussian spread of the semiclassical state, where $\rho$ is the energy density of the scalar field,  $\rho_\star =  (8 \pi G  \beta^2 \Delta/3)^{-1}$, and $\Delta$ is the (constant) area gap.  The maximum value of the energy density (reached at the bounce) is $\rho_{\text{max}} = \sqrt{\rho_\star^{3} /\rho_0}$, where $\rho_0 = \rho\big|_{a = 1}$.

Comparing Eq.~\eqref{eq:friedmann_2vertex_R} with $\kappa=0$ (as corresponds to vanishing spatial curvature) and~\eqref{eq:effective_friedmann_LQC} we observe that the U($N$)-symmetric two-vertex model describes the old effective dynamics of LQC with an area gap   $\Delta=1/(2\gamma)^2$ and a minimum fiducial length $\mu_0 = \sqrt{\Delta}=1/(2\gamma)$.

\subsection{Towards LQC improved dynamics}

The U($N$)-symmetric two-vertex model describes the  old effective dynamics of LQC. Nevertheless, we can try to implement the improved dynamics~\cite{Ashtekar_Pawlowski_ImprovedDynamics2006} in a similar way to that of LQC by suitably modifying the model. 

To do so, let us go back to the relation~\eqref{eq:espinores_fase_U(N)} imposed by the U($N$) symmetry on the spinors. Note that, in the U($N$) sector, the Poisson structure~\eqref{eq:commutation} for the spinors  determines the Poisson bracket between the twist angle $\varphi$ and the total area $A$, so that they are canonically conjugate variables. To introduce the improved dynamics, we need to change this structure so that $\varphi$ and $A$ are no longer canonically conjugate. One way of doing this is to change the Poisson bracket to $\{\varphi,2\sqrt A\}=1$, while keeping the Hamiltonian~\eqref{eq:hamiltonian_UN_generalisation}. 
Another equivalent option, which is the one we will use in the following, is to introduce a new variable $\phi=\varphi/\sqrt A$ canonically conjugate to $A$, so that $\{\phi,A\}=1$. Then, the U($N$) relation~\eqref{eq:espinores_fase_U(N)} now becomes 
\begin{equation}
\label{eq:espinores_fase_U(N)_improved}
    |w_i\rangle =e^{-i\phi\sqrt A / 2}|\bar{z}_i\rangle.
\end{equation}
The spinor Poisson structure that leads to this modification is no longer~\eqref{eq:commutation}, so this modified model is not a direct consequence of the two-vertex model but of a suitably modified one, whose structure is beyond the scope of this paper. 
The Hamiltonian in terms of these new canonical variables $\phi$ and $A$ is  
\begin{equation}
\label{eq:hamiltonian_UN_improved}
    \Hredg = 2 \mathcal M  \area^2 \left[\gamma \left(-1  +   \cos \big(\sqrt{A}  \phi \big) \right) + g(\area) \right],
\end{equation}
where we have set $\kappa=0$ as before for comparison with LQC.

If we now follow  steps analogous to those of Sec.~\ref{sec:LQC_effective_dynamics}, then we obtain 
\begin{equation}
\label{eq:2_vertex_friedmann_improved}
    \frac{1}{N^2} \frac{\dot{a}^2}{a^2} = \gamma f(a) \Big[ 1 - \frac{\beta^3}{4\gamma} f(a)\Big],
\end{equation}
where now $f(a)=  2 \beta^{-3}g(a^2/\beta)$.

On the other hand, the improved dynamics formalism introduces in the holonomies a nonconstant minimum fiducial length $\bar{\mu} = \sqrt{\Delta}/a$. The resulting effective LQC equation is~\cite{Taveras_CorrectionsFriedmann2008}
\begin{equation}
\label{eq:effective_friedmann_LQC_taveras}
    \frac{1}{\mathcal{N}^2} \frac{\dot{a}^2}{a^2} = \frac{8 \pi G}{3} \rho \Big(1 - \frac{\rho}{\rho_{\star}} \Big).
\end{equation}
Therefore, now the maximum energy density (reached at the bounce) is precisely $\rho_{\star}$ and, thus, independent of the initial energy density $\rho_0$. We  see that the modified U($N$)-symmetric two-vertex dynamics~\eqref{eq:2_vertex_friedmann_improved} provides the effective improved dynamics~\eqref{eq:effective_friedmann_LQC_taveras} with energy density $\rho = f/\mathcal{V}_{\Sigma}$, gravitational constant $G=3\gamma \mathcal V_\Sigma/(8\pi)$ and area gap $\Delta = \beta/(4\gamma^2)$. Consequently, the $\bar \mu$ of the LQC improved dynamics can be written in terms of the modified two-vertex parameters as $\bar\mu=\sqrt\Delta/a=\sqrt\beta/(2\gamma a)$.

As we have already mentioned, this result is a consequence of an \textit{ad hoc} modification of the two-vertex model (very much in the same way as the improved dynamics is introduced in LQC). Nevertheless, this modification is useful to further understand the underlying physics behind the imposition of the improved dynamics in LQC and its relation with full LQG.

 \section{Conclusions}
\label{sec:conclusions}

The two-vertex graph has proved to be a useful model within LQG in order to implement dynamics and to study the emergence of an interesting cosmological behavior, at least in a symmetry-reduced sector~\cite{Borja_Freidel_Garay_Livine_2010_Return, Borja_Diaz-Polo_Garay_Livine_2010_Dynamics, Borja:2011ha, Aranguren_Garay_Livine_2022,Livine_Martin-Benito_ClassicalSettingEffective2013}. Therefore, this model provides an excellent arena to explore the main open problems of the theory. 
On the other hand, the spinorial formalism introduced in~\cite{Livine2011} gives us a convenient mathematical formulation of the kinematic Hilbert space of LQG.

We have made use of the spinorial formalism applied to the two-vertex model in order to analytically study  the evolution of the classical volume. This formalism provides normal vectors to the faces of the closed polyhedra (the conditions for the Minkowski theorem are constraints) associated with each of the vertices. 
Nevertheless, there is no known analytical formula to compute the volume of a general polyhedron (with $N>4$ faces) in terms of the normal vectors that generate it. 
Instead, we have used geometric quadrupoles~\cite{goellerProbingShapeQuantum2018} in order to  construct an approximation of the volume that reproduces the qualitative behavior of the volume~\cite{Aranguren_Garay_Livine_2022}.
We have calculated the equation of motion for this approximation, finding that degenerate polyhedra---with all the faces parallel to each other---have null volume forever, i.e., zero-volume  singularities are decoupled from the rest of configurations. 

The $\text{U}(N)$-symmetric sector of the two-vertex model contains just one degree of freedom corresponding to the twist angle and the total area of the polyhedra, which are canonically conjugate. We have found that the approximation to the volume in terms of quadrupoles is proportional to the exact volume for the case of the tetrahedron ($N=4$). It is left for future work to check whether this behavior is true for any other number of faces.
Comparing the evolution of the quadrupole approximation to the volume and the  total area in the $\text{U}(N)$-symmetric sector, we have found that they precisely follow the evolution one expects from a homogeneous and isotropic evolution. This relation is yet another argument for the use of the quadrupole approximation of the volume and for the interpretation of the  two-vertex model as a cosmological model (at least in the $\text{U}(N)$-reduced sector).

We have shown that a generalization of the two-vertex Hamiltonian constraint (that preserves the symmetries of the model) consisting of the addition of a function of the area, effectively takes into account matter content in the form of arbitrary barotropic perfect fluids. In particular, we have shown that there is a direct relation between the equation of state and the specific function of the area introduced in the two-vertex Hamiltonian.

Finally, we have proved that the two-vertex model is indeed an LQG truncation that effectively describes LQC with arbitrary barotropic perfect fluids. The LQC flavor that derives directly from the two-vertex model is the old dynamics. However, a suitable modification of the Poisson bracket structure so that the twist angle is canonically conjugate to the square root of the area (and not the area itself) leads to the LQC improved dynamics.

\acknowledgments
We would like to thank Etera R. Livine and Mercedes Martín-Benito for very enlightening conversations. This work was partially supported by Grants No. PID2020–118159GB-C44, No. PID2021-123226NB-I00 and No. PID2022-139841NB-I00 (funded by MCIN/AEI/10.13039/501100011033 and by “ERDF A way of making Europe”), and the Basque Government Grant No. IT1628-22. Additionally, Á. C. acknowledges financial support from MIU (Ministerio de Universidades, Spain) fellowship FPU22/03222 and IPARCOS Ayudas de Máster IPARCOS-UCM/2022.
\bibliography{bibliography}

\end{document}